\documentstyle[aps,preprint,eqsecnum,epsfig,floats]{revtex}
\tighten
\begin{document}
\draft
\preprint{\parbox{6cm}{\flushright UAB--FT--456\\IFT--P.078/98\\October
    1998\\[1cm]}}

\vspace{1in}
\title{\Large \bf Higgs-- and Goldstone bosons--mediated\\
long range forces}

\author{ {\bf F. Ferrer}
\thanks{fferrer@ifae.es} 
}                   
\address{ Grup de F\'{\i}sica Te\`orica and Institut de F\'{\i}sica d'Altes
  Energies\\Universitat Aut\`onoma de Barcelona\\ 
08193 Bellaterra, Barcelona, Spain} 

\author{{\bf M. Nowakowski}
\thanks{marek@ift.unesp.br}
}
\address{Instituto de F\'{\i}sica Te\'{o}rica, Universidade Estadual
  Paulista\\Rua Pamplona 145, 01405-900 S\~ao Paulo, Brazil} 
\maketitle
\vfill
\begin{abstract} 
In certain mild extensions of the Standard Model, spin-independent
long range forces can arise by exchange of two very light 
pseudoscalar spin--$0$ bosons. In particular, we have in mind  
models in which these bosons do not have direct tree level couplings to 
ordinary fermions. Using the dispersion theoretical method, we find a
$1/r^{3}$ behaviour of the potential for the exchange of very light
pseudoscalars and a $1/r^{7}$ dependence if the pseudoscalars are true
massless Goldstone bosons. 
\end{abstract}                                                                
\pacs{}

\section{Introduction}
Most studies and investigations on long range forces have always
centered, for obvious reasons, around the electromagnetic
and gravitational interaction. However, starting with the very early 
example of the Casimir--Polder long range force \cite{casimir}, 
over the Feinberg--Sucher force \cite{FS}
mediated by two neutrinos~(see Fig.1) 
and finally going to recent developments in 
supersymmetry and superstrings \cite{string}, 
there has been a continuous interest in effects
and detection of exotic long range forces~\cite{meu}. The actual applicability or
relevance of these forces is, of course, different from case to case.
For instance, the Casimir--Polder force is, in principle, of electromagnetic 
origin. It arises as a consequence of photon exchange between polarizable
neutral systems and the resulting potential has a $1/r^7$ dependence at long
distances. Although the Casimir--Polder force has been recently detected in a laboratory experiment \cite{casimir.exp}, 
the neutrino mediated (i.e. involving weak interaction couplings) 
Feinberg--Sucher force is too weak to be of any significance in Earth--based
experiments. If at all, a suitable arena for this force would be of 
astrophysical and/or cosmological dimension (see for instance in this respect
\cite{hartle} and references in \cite{fgn}). 
The result of Feinberg and Sucher
has been recently extended to also account for the exchange of very light 
Dirac \cite{fisch} and Majorana \cite{majorana}
neutrinos. Temperature--dependent corrections including the exchange 
of thermalized neutrinos at finite temperature, like the relic cosmic 
neutrinos at $T^{-1} \sim 1 mm$, have been calculated in \cite{HP,fgn}.
Finally let us mention that extensions of the Standard Model 
can allow, in principle, for a variety of different long range
forces~\cite{meu}, mediated, for instance by very light or massless scalars or
pseudoscalars~\cite{meu2}. 
The former force acting between neutrinos themselves 
has been discussed e.g. in \cite{moha}. The potential due to the exchange of
two pseudoscalar particles (box--diagrams) was computed in \cite{box,tes}.
Furthermore, new exotic long range forces can 
appear also in the context of gauge
mediated supersymmetry breaking  and in superstring theories \cite{string}.
The implications of a new long range force due to an extra $U(1)$ gauge group
have been discussed recently by Fayet in \cite{fayet}.
\begin{figure}[bht]
\begin{center}
\epsfig{file=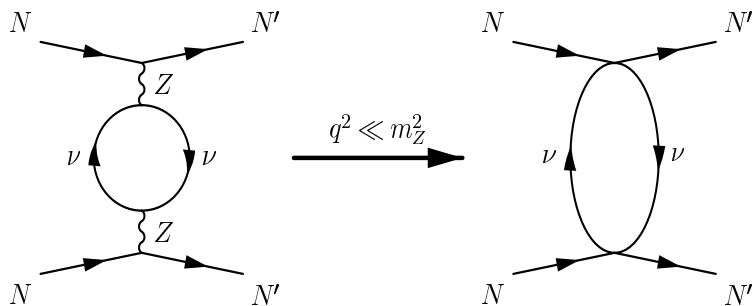,width=10cm,height=4cm}
\end{center}
\caption{{\it One of the diagrams in the S.M. giving rise to the two
    neutrino force in four-Fermi effective theory.}}
\end{figure}

Should neutrinos be the only massless or very light particles in 
spectrum of the theory, then the Feinberg--Sucher result would be the only
possible exotic 
long range force, regardless of the model. This is clear since all
long range forces (including the electromagnetic and gravitational interaction)
arise as a consequence of an exchange of very light quanta. However, as 
mentioned above many extensions of the Standard Model also predict very light
pseudoscalars. Usually diagrams involving two such pseudoscalars
will then result in a spin--independent 
long range force between standard fermions~\cite{box,tes}. 
Recall that a exchange of a single pseudoscalar between fermions gives
a spin--dependent result for the potential~\cite{meu}. Indeed, a covariant calculation 
with two pseudoscalars exchange has been
recently performed in \cite{tes} in the context of a generic theory 
where the coupling of the pseudoscalar $\phi$ to fermions is taken either
as $\phi \bar{\psi}\gamma_5 \psi$ 
or, alternatively, as the derivative 
version
$(\partial_{\mu}\phi)\bar{\psi}\gamma_5 \gamma^{\mu}\psi$, in the latter case
$\phi$ can represent a generic Goldstone boson. In the first case the authors
obtain the a $1/r^{3}$ dependence of the potential whereas the double exchange 
of Goldstone bosons yields a more drastic fall--off, viz $1/r^{5}$.
However, very often, i.e. in a wide class of models, these pseudoscalars
do not couple to standard fermions (often not even to gauge bosons)
 on account of some symmetry arguments (see appendix where one such model is
 briefly sketched). However they do 
always have a tree level coupling to Higgs--scalar particles of the theory.
Indeed, it is difficult to imagine a reasonable symmetry 
argument which would forbid such couplings. We now assume that the scalars
themselves
couple to standard fermions, which is the case in most models. If so,
then the diagram in Fig. 2 displays a very nice analogy to the diagram 
responsible for the Feinberg--Sucher force (see Fig. 1). Indeed, we have
replaced only fermions by bosons when comparing Fig. 1 with Fig. 2.
Of course, one expects a different $r$ dependence of the potential arising
from the two diagrams due to different
dimensionality of the the coupling constants. 
If the pseudoscalars have both couplings, to the fermions
as well as to the Higgses, the result of 
\cite{tes} and our paper should then
be added.
Since the coupling of the
scalar Higgs to fermions is usually proportional to the mass of the fermion,
one may suspect that the box--diagrams using the direct pseudoscalar--fermion
coupling are more important. In general this is model--dependent,
but we can safely state here that the pseudoscalar fermion coupling constant
is also `experimentally' restricted by arguments of energy loss in stars
where one assumes that the bulk of energy of the star is carried away
by the standard mechanism in form of photons and neutrinos \cite{raffelt}. 

\begin{figure}[bht]
\begin{center}
\epsfig{file=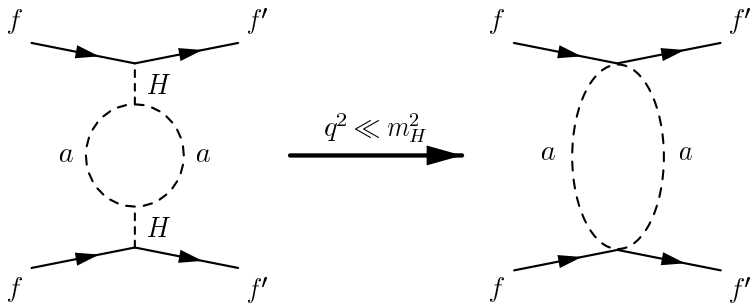,width=10cm,height=4cm}
\end{center}
\caption{{\it Pseudoscalar mediated long range force without direct
    fermion coupling.}}
\end{figure}

If we assume that the pseudoscalar is a Goldstone boson, a connection
to the $U(1)$--forces considered in \cite{fayet} can be possibly made as the
latter display a `Goldstone--like' behaviour as we approach with the $U(1)$  
coupling to zero \cite{fayet}.

The paper is organized as follows. In section 2 we calculate, using
dispersion theoretical methods~\cite{fs.review}, the long range
force due to the diagram in Fig. 2 where we assume that the coupling between
the Higgs ($H$) and the very light pseudoscalar ($a$) is linear and of the 
form $Haa$. We also briefly touch some issues concerning 
a possible temperature dependence of the potential.  
In the subsequent section we change the linear coupling to a derivative
version of the form $H(\partial_{\mu}a)(\partial^{\mu}a)$. In section 4
we discuss the particular case of Goldstone bosons exchange. In section 5
we summarize our results.

\section{Long range forces due to pseudoscalar-pseudoscalar-scalar
non-derivative couplings}
The dispersion theoretical technique of calculating long range forces in
quantum field theory is reviewed in detail in \cite{fs.review}. 
This method is especially suitable to cope with higher order diagrams and
relativistic effects and its implementation to compute the neutrino pair
exchange force is straightforward~\cite{FS}. The results agree with the
computations done in~\cite{hsu} by performing the Fourier transform of the
associated Feynman amplitude in momentum space~(this latter strategy is only
applicable in general when there is no lower order long range force and
relativistic corrections are negligible).

According to  the rules of the dispersion theoretical method we must compute
the following Laplace transform (we restrict here ourselves to central forces which depend only on the distance $r \equiv |{\mathbf r}|$
between the two particles):
\begin{equation} \label{defpot}
V(r)={-i \over 8\pi^2 r}\;\int^{\infty}_{4m_a^2}{\!
dt \: [{\cal M}]_t\,\exp (-\sqrt{t}\,r)}
\end{equation}
where the integration variable $t$ stands for the usual Mandelstam variable
which equals the four--momentum transfer
squared, $q^2$. Here, 
$[{\cal M}]_t$ denotes the discontinuity
of the Feynman amplitude (i.e. the absorptive part of the same) across the cut
in the real $t$ axis and is best computed by
taking advantage of the analyticity and generalized unitarity properties
leading to the Cutkosky rules~\cite{fs.review}.

Let us now consider the case of some generic interaction terms of the form
\begin{equation} \label{int.generic}
{\cal L}_{int}=g_{{}_{Hff}}\bar{f}fH, \,\,\,\,\, 
{\cal L}'_{int}=g_{{}_{Haa}}aaH
\end{equation}
where $f$ are standard fermions, $H$ is the heavy Higgs with mass
$m_H$ and $a$ is the very light pseudoscalar with mass $m_a$. We can
essentially neglect here possible quartic couplings of the form $H^2a^2$
as self--energy corrections due to this quartic coupling would only eventually
give rise to contact interactions. 

It is convenient to define global coupling constants as
\begin{equation} \label{GG'}
G \equiv {g_{{}_{Hff}}g_{{}_{Haa}} \over m_H^2}, \,\,\,\,\,
G' \equiv {g_{{}_{Hf'f'}}g_{{}_{Haa}} \over m_H^2}
\end{equation}
which capture the constants of the four vertices and the two Higgs
propagators in Fig.2. For future reference we draw the reader's attention
to the fact that we have expanded the Higgs propagators in $q^2$ and
kept only the zeroth order of this expansion; this then gives the $m_H^2$ in
the denominators of $G$ and $G'$ in (\ref{GG'}).  
The full matrix element of the diagram in Fig.2
is given by
\begin{equation} \label{M}
{\cal M}=-2\, i\, G G' \: \Gamma \:\left[\bar{u}(p_1')u(p_1)\; \bar{u}(p_2')u(p_2)
\right] .
\end{equation}
The one--loop integral is represented above by $\Gamma$ i.e.
\begin{eqnarray} \label{Gamma}
\Gamma &\equiv& \int {\!\! d^4k \over (2\pi)^4}{i \over k^2 -m_a^2
  +i\epsilon} \,{i \over \bar{k}^2 -m_a^2 +i\epsilon} \nonumber \\
\bar{k}=k-q,&\quad& \quad q=p_1-p_1'=p_2'-p_2, \quad \quad q^2=t\:.
\end{eqnarray}
We assume also the non--relativistic limit in which we have
$\bar{u}(p_1')u(p_1)=\bar{u}(p_2')u(p_2)\simeq 1$. Using the prescriptions
arising from generalized unitarity, which amount to the replacement,
\begin{equation}
\frac{1}{k^2-m_a^2 +i\epsilon}\;\; 
\longrightarrow -2 \pi i \: \delta (k^2-m_a^2) \: \Theta (k^0),
\end{equation}
we obtain for the discontinuity 
\begin{eqnarray} \label{calc.Gamma}
[\Gamma]_t&=&\frac{1}{(2\pi)^2} \int {\!\!{d^4 k \over (2 \pi)^4}\:
\delta(k^2-m_a^2)\: \delta(\bar{k}^2-m_a^2)\: \theta(k^0)\:\theta(\bar{k}^0)}\nonumber
\\
&=& {1 \over 8\pi}\: \sqrt{1-\frac{4 m_a^2}{t}}.
\end{eqnarray}
Obviously we have $[{\cal M}]_t=-2\,i\,GG'\:[\Gamma]_t$ which has to be inserted
into (\ref{defpot}) to compute the final expression of the potential:
\begin{eqnarray} \label{finalpot}
V(r)&=& -\frac{GG'}{32 \pi^3 r}\;\int^{\infty}_{4m_a^2}{\!dt \: \sqrt{1 -
\frac{4m_a^2}{t}}\; \exp (-\sqrt{t}\, r)}
\nonumber \\
&=& -\frac{GG' m_a}{8 \pi^3 r^2} \; K_1(2m_ar)
\end{eqnarray}
where $K_1$ is a modified Bessel function. To show that (\ref{finalpot}),
for a very small mass $m_a$, yields indeed a long range potential, let us 
take the limit $m_a \to 0$ in (\ref{finalpot}) (equivalent to 
$rm_a << 1$). 
For the leading order
of the expansion we get
\begin{equation} \label{zeromass}
V(r) \simeq -\frac{GG'}{16 \pi^3 r^3}.
\end{equation}
For comparison we quote below the Feinberg--Sucher result for massless
neutrinos \cite{FS}
\begin{equation} \label{FS}
V_{FS}(r) = {G^2_F g_v g_{v'} \over 4 \pi^3 r^5}
\end{equation}
where $G_F$ is the Fermi and $g_v$ and $g_{v'}$ weak vector coupling constants.
Note that, in contrast to (\ref{zeromass}), the Feinberg--Sucher force
(\ref{FS}) is repulsive. This difference between these two forces is due to an
extra minus sign for the fermion--loop in (\ref{FS}).

We would like to touch at this point briefly upon finite temperature 
corrections to (\ref{finalpot}) and (\ref{zeromass}). In doing so we
will follow mainly \cite{HP} and \cite{fgn} to which we refer the reader for
more details on this subject. At finite temperature $T$ the spin--$0$ boson
propagator $S_T(k)$ takes the form
\begin{equation} \label{prop}
S_T(k)={1 \over k^2 -m_a^2 +i\epsilon} 
-2i\pi \: \delta (k^2 -m_a^2) \: n(T),  
\end{equation}
where $n(T)$ is the particle distribution function with the chemical
potential already set to zero.
As noted explicitly in \cite{HP}, the propagator (\ref{prop}) is sufficient
to calculate the problem at hand\footnote{We depart for a moment from the
  dispersion theoretical method and use, following~\cite{HP} and~\cite{fgn},
  the traditional Fourier transform to compute the $T$--dependent
  effects. In such a situation we need only the real part of the amplitude
  correctly given by using~(\ref{prop}), (see ref.~\cite{HP}), which is the $1-1$ component of the
  full $2$-dimensional matrix propagator used in the real time approach to
  finite temperature field theory~\cite{temp}.}.

We will restrict ourselves to Boltzmann distributions, 
\begin{equation} \label{distrib}
n(T)=\exp \left[(-E_k)/T\right],
\end{equation}
in which $E_k$ is the energy. 
To calculate
the potential itself we use now the method of Fourier--transforming the
momentum amplitude i.e.
\begin{equation} \label{potdef2}
V_T(r) = \int{\!\! {d^3 {\mathbf Q} \over (2\pi)^3} \; \exp(i{\mathbf
    Q}{\mathbf r})\:{\cal M}_T ({\mathbf Q})}={1 \over 2 \pi^2
    r}\int^{\infty}_0 \!dQ\: Q \: {\cal M}_T(Q)\: \sin Qr
\end{equation}
where in the static limit we have $q \simeq (0,{\mathbf Q})$ and in the second
equality we have defined $Q=|{\mathbf Q}|$ and $r=|{\mathbf r}|$. The second expression
in (\ref{potdef2}) holds for potentials which depend only on $r$.
As before, we can write effectively ${\cal M}_T \simeq -2i GG' \: \Gamma_T$
such that $\Gamma_T$ is the one loop--integral involving 
two `cross' products of two
propagators, one the standard vacuum part and the other thermal part, viz
\begin{equation} \label{gamma}
\Gamma_T= \int{\!\! {d^4 k \over (2\pi)^4}\: 2i\pi \:\delta (k^2-m_a^2) \: n(T)
\left({1 \over (k+q)^2-m_a^2} + {1 \over (k-q)^2 -m_a^2}\right)}.
\end{equation}
$\Gamma_T$ can be further evaluated to be
\begin{equation} \label{gamma2}
\Gamma_T={4i \over (2\pi)^2}\:\int_0^{\infty}{\!\!{dk k^2 \over \sqrt{k^2 +m_a^2}}
\exp(-\sqrt{k^2 +m_a^2}/T)\int_{-1}^1 {dz {1 \over 4k^2z^2 -Q^2}}}
\end{equation}
where now $k=|{\mathbf k}|$. Recalling that ${\cal M}_T=-2iGG'\:\Gamma_T$
and inserting this into (\ref{potdef2}) and subsequently performing the 
integration first over $Q$ and then over $z$ we get
\begin{eqnarray} \label{potT1}
V_T(r) &=& -{GG' \over 4\pi^3}{1 \over r^2}\:\int_0^{\infty}{
\!{k\,dk \over \sqrt{k^2
+m_a^2}} \exp \left(\sqrt{k^2 +m_a^2}/T \right)\: \sin (2kr)} \nonumber \\
&=&- {GG' \over 2 \pi^2}{1 \over r}{Tm_a \over \sqrt{1 + (2rT)^2}}
\; K_1\! \left( \displaystyle{{ m_a \over T}}\sqrt{1 +(2rT)^2} \right).
\end{eqnarray}
Equation (\ref{potT1}) is the finite temperature correction to 
(\ref{finalpot}). It is instructive at this stage to examine different limits of
(\ref{potT1}). First, let us go with the mass $m_a$ to zero as done in
(\ref{zeromass}) for the
vacuum contribution. We get the simple result
\begin{equation} \label{zeromass2}
V_T(r) \simeq -{GG' \over 2 \pi^3}{1 \over r}
{T^2 \over 1 + (2rT)^2}.
\end{equation}
Using the last limit (i.e. $m_a \to 0$) we can also investigate
the range $r \gg T^{-1}$. In this range (\ref{zeromass2}) can be
expanded to give 
\begin{equation} \label{zeromass3}
V_T(r) \simeq -{GG' \over 8\pi^3 r^3}.
\end{equation}
Remaining in this long distance range, as compared to the temperature
inverse, we can add now to the vacuum part (\ref{zeromass}) equation 
(\ref{zeromass3}) to arrive at the complete answer for the potential
\begin{equation} \label{tot}
V_{tot}(r)=V_T(r)+V(r) \simeq -{3 \over 16}
{GG' \over \pi^3 r^3}
\end{equation}
This last result is particularly interesting when we compare it with
the corresponding result in the context of the two neutrino force,
calculated at zero and finite temperature \cite{fgn}. 
In the 
neutrino 
case the total sum consisting 
of the vacuum part and the finite 
temperature contribution (i.e. an equation corresponding to (\ref{tot}))
switches the sign of the force in the range $r \gg T^{-1}$, a repulsive force
becomes attractive in the presence of relic neutrinos \cite{fgn}. This is a
quite interesting result which sheds new light on the Feinberg--Sucher
force. 
The reason why a similar
reversal does not take place in the two boson force (cf. eq.(\ref{tot}))
(i.e.
why this   
attractive
force does not become repulsive when we add temperature corrections)
is due to the fact that the relative sign between the vacuum part of the 
propagator and the thermal part is plus in the boson propagator (cf. eq.
(\ref{prop})) whereas it is minus for fermions \cite{temp}.

Although the temperature of the very light pseudoscalars 
at the present epoch, provided of course these pseudoscalars exist, 
is model dependent, it should be comparable 
(at least in the order of magnitude) 
to the temperature of relic Axions \cite{kolb} or Majorons \cite{babu}.

\section{The case of derivative couplings}
In this section we will also compute the dispersion force arising from Fig.2,
considering however a different coupling scheme between the heavy Higgs and the
light pseudoscalars. For the relevant Lagrangian interaction we take now
\cite{suzuki}
\begin{equation} \label{interaction2}
{\cal L}''_{int}=\tilde{g}_{{}_{Haa}}H(\partial^{\mu}a)(\partial_{\mu}a).
\end{equation}
To simplify things, we will also start right from the beginning considering
massless pseudoscalars (instead of taking the limit
$m_a \to 0$ at the end of the calculation). We define also
over--all couplings in analogy to (\ref{GG'})
\begin{equation} \label{couplings2}
\tilde{G} \equiv {g_{{}_{Hff}}\tilde{g}_{{}_{Haa}} \over m_H^2}, \,\,\,\,\,
\tilde{G}' \equiv {g_{{}_{Hf'f'}}\tilde{g}_{{}_{Haa}} \over m_H^2}.
\end{equation}
As in the preceding section we start with the dispersion 
theoretical definition
of the potential i.e. eq. (\ref{defpot}) where we denote now the matrix
element by $\tilde{{\cal M}}$ given by
\begin{eqnarray} \label{matrixelement2}
\tilde{{\cal M}}&\simeq&  -2 i \tilde{G}\tilde{G}'\cdot \tilde{\Gamma} 
\nonumber \\
\tilde{\Gamma}&=&\int{\!\!\frac{d^4k}{(2\pi)^4}\,\frac{i}{k^2}\,\frac{i}{\bar{k
      }^2}\,(k \cdot \bar{k})^2}
\end{eqnarray}
where as before $\bar{k}=q-k$. The rest of the calculation follows essentially
the same line as in section 2. First we have to calculate the discontinuity
$[\tilde{{\cal M}}]_t \propto [\tilde{\Gamma}]_t$ and insert the result into
equation (\ref{defpot}). For the discontinuity we obtain
\begin{eqnarray} \label{discont2}
\left[ \tilde{\Gamma} \right]_t&=&\frac{q^{\mu}q^{\nu}}{(2\pi)^2} \int { d^4k
  \;  \delta(k^2) \delta(\bar{k}^2)\: k_{\mu}k_{\nu}} \nonumber \\
&=&\frac{q^{\mu}q^{\nu}}{(2\pi)^2}\frac{\pi}{2}\left[
\frac{1}{3} \left(q_{\mu}q_{\nu}-\frac{1}{4}g_{\mu \nu} q^2 \right)
  \right]\nonumber\\ 
&=&\frac{t^2}{32 \pi}
\end{eqnarray} 
with $q^2=t$ as usual.
It remains calculating the integral transform of this discontinuity.
To distinguish the potential from the results in the
preceding section we will call the
potential due to two pseudoscalar exchange arising from the interaction
(\ref{interaction2}), $\tilde{V}$. For the latter we get
\begin{eqnarray} \label{potderivative}
\tilde{V}(r)&=&-\frac{\tilde{G}\tilde{G}'}{128 \pi^3 r} \: \int_0^{\infty}
{ \!\! dt  \: \exp (-\sqrt{t}\:r) \: t^2} \nonumber \\
&=& -\frac{15\tilde{G}\tilde{G}'}{8 \pi^3 r^7}.
\end{eqnarray}
If we compare this expression with the potential (\ref{zeromass}) it becomes
clear that it is the $q^4=t^2$ dependence of $[\Gamma]_t$ which gives here the
steep fall--off proportional to $1/r^7$. In (\ref{zeromass}) the corresponding
integrand i.e. $[\Gamma]_t$ was simply a constant (for $m_a=0$)
giving rise to a milder $1/r^3$ dependence.

In principle, one could now also calculate temperature--dependent 
effects as we have done in section 2. We will, however, not dwell further
on this subject here and instead address in the next section the interesting
question of the potential due to the exchange of two Goldstone bosons.

\section{Long range forces due to physical Goldstone bosons}
In the two preceding sections we have calculated in a rather 
model--independent
way the potential due to two pseudoscalar exchange according to
Fig.2 and using two different interaction lagrangians, (\ref{int.generic})
and (\ref{interaction2}). Here we would like to address the situation when
the pseudoscalar is a true (i.e. strictly massless) Goldstone
boson. 

In the literature one can find numerous papers where for Goldstone
bosons either the linear scheme (\ref{int.generic}) is used
or the derivative one as in (\ref{interaction2}), 
very often with the insistence
that, for Goldstone bosons, the derivative coupling is the correct one.

We will examine the two Goldstone bosons potential not in a general model,
but using as an example the singlet Majoron model \cite{singlet}, briefly
sketched in the
appendix. The Majoron $J$ (we change the notation here, $a \to J$) is a 
true Goldstone boson due to spontaneous breaking of the lepton number.
The two different couplings discussed above have been derived explicitly
in the appendix. Equation (\ref{int1}) corresponds to the linear scheme whereas
(\ref{int2}) to the derivative one. Also note that, apart from the explicit
form of the couplings, we can use now on the results from the two preceding
sections.

Since in the singlet Majoron model the physical spectrum consists of two {\it
  heavy} scalars $H$ and $S$ and the {\it massless} Majoron $J$, instead of
  one diagram as in Fig.2, we have four distinct amplitudes corresponding to
  the four possible combinations of the heavy scalars i.e. to the exchange
  $HH$, $SS$, $HS$ and $SH$.  

Let us first investigate in detail the linear coupling scheme (\ref{int1})
which would then fall in the general domain of section 2. 
All we have to do now is to use the result (\ref{finalpot}) and replace the
general coupling
$GG'$ by the concrete example from the appendix. 
As mentioned before, we have to
sum over the different possibilities of heavy scalar exchanges i.e.
\begin{equation} \label{GGps}
(GG')_{{}_{{\mathrm Majoron}}}
=\sum_{P,P'=H,S}{{g_{{}_{Pff}}g_{{}_{P'ff}} g_{{}_{PJJ}}
g_{{}_{P'JJ}} \over m_P^2 m_{P'}^2}}.
\end{equation}
Although the coupling of scalar Higgses is not always strictly proportional
to the fermion mass (for instance, in case of nucleons it also depends on the
gluon content of the nucleons) we will use here, as an example, 
the coupling of $H$ and $S$ to fundamental fermions. In the singlet
Majoron model they are given by
$g_{{}_{Hff}}=-i (\sqrt{2} G_F)^{1/2} m_f \cos \theta$ and 
$g_{{}_{Sff}}=-i(\sqrt{2}
G_F)^{1/2} m_f \sin \theta$. The coupling constants among the spin--$0$ bosons
can be read off from~(\ref{int1}). Taking all this into account we obtain
\begin{equation} \label{GGps0}
(GG')_{{}_{{\mathrm Majoron}}}=0.
\end{equation}
This, of course, does not imply that the potential due to the exchange
of two Majorons is zero. It means, however, that it is not of the simple
$1/r^3$ dependence as indicated in (\ref{zeromass}). In order to get
a meaningful non-zero result for the potential (due to Majorons), we have to 
go one step more in the $q^2$--expansion of the heavy Higgs propagators.
We already stressed in section 2 that the results presented there
are valid for the zeroth order expansion i.e. fully neglecting 
the $q^2$ in the heavy Higgs
propagators. In other words, this means that   
$(GG')_{{}_{{\mathrm Majoron}}}=(GG')_{{}_{{\mathrm Majoron}}}(q^2=0)=0$.
The next term in the expansion is
\begin{eqnarray} \label{GGpsnon0}
(GG')_{{}_{{\mathrm Majoron}}}(q^2)&\equiv &
\sum_{P,P'=H,S}{{g_{{}_{Pff}}g_{{}_{P'ff}}
g_{{}_{PJJ}}g_{{}_{P'JJ}} \over (q^2 - m_P^2)(q^2 -  m_{P'}^2)}}
\nonumber \\
& \simeq & \frac{G_F^2 m_f m_{f'}}{2}\sin^2 \theta
\cos^2 \theta \tan^2 \beta \left(\frac{1}{m_H^2}-\frac{1}{m_S^2}\right)^2
q^4\;.
\end{eqnarray}
Since the relevant integrand in the form of $[\Gamma]_t|_{m_a=0}$ (cf.eq. 
(\ref{finalpot})) does not give any further $q^2$ dependence
(it is a constant), the $q^4=t^2$ term
from (\ref{GGpsnon0}) is the only one to be integrated over. This, of course,
resembles the $q^4$ dependence in (\ref{discont2}). 
Indeed, the final expression
for the potential reads
\begin{equation} \label{finalpot2}
V_{JJ}(r)=-\frac{15 G_f^2 m_f m_{f'}}{16 \pi^3 r^7}\sin^2(2\theta)
\tan^2 \beta \left(\frac{1}{m_H^2}-\frac{1}{m_S^2}\right)^2 
\end{equation}
and has remarkably the same $r$--dependence as (\ref{potderivative}).

Let us now repeat the steps from above for the derivative coupling scheme 
(\ref{interaction2}) discussed in the general setting in section 3 and given
specifically for the singlet Majoron case in (\ref{int2}). The equation
corresponding to (\ref{GGpsnon0}) reads in this scenario as follows
\begin{eqnarray} \label{GGderiv}
(\tilde{G}\tilde{G}')_{{}_{{\mathrm Majoron}}}(q^2)&\equiv &
\sum_{P,P'=H,S}{{g_{{}_{Pff}}g_{{}_{P'ff}}
\tilde{g}_{{}_{PJJ}}
\tilde{g}_{{}_{P'JJ}} \over (q^2 - m_P^2)(q^2 -  m_{P'}^2)}}
\nonumber \\
& \simeq &\frac{G_F^2 m_f m_f'}{2}\sin^2 2\theta
\tan^2 \beta \left(\frac{1}{m_H^2}-\frac{1}{m_S^2}\right)^2  +...
\nonumber \\
&\simeq &
(\tilde{G}\tilde{G}')_{{}_{{\mathrm Majoron}}}(q^2=0)
\end{eqnarray} 
i.e. a non--zero result of the expansion here is already possible at the
lowest order. Inserting this into eq. (\ref{potderivative}) 
we confirm, however,
the result (\ref{finalpot2}). This is mainly due to the fact that 
$(GG')_{{}_{{\mathrm Majoron}}}(q^2)$ has the same $q^2$
dependence as $[\tilde{\Gamma}]_t$.

The equivalence of the two coupling schemes, 
(\ref{int1}) and (\ref{int2}), in calculating the potential
due to Majorons exchange is a particular example of a more general theorem
which states that physical results cannot depend on the 
chosen parametrisation of the fields \cite{theorem}.
Recall that (\ref{int1}) follows directly from choosing the representation
(\ref{rep1}) whereas (\ref{int2}) is a consequence of the 
representation (\ref{rep2}).

\section{Conclusions}
We have calculated the long range potentials due to the exchange
of very light or massless pseudoscalars using dispersion theoretical
methods. 
In particular, we investigated
these long range potentials in models where the very light pseudoscalars
do not have a tree-level coupling to the standard fermion. The only possible
diagram which in coordinate space can then result in long range potentials
displays a formal resemblance to the diagram responsible for the two 
neutrino Feinberg--Sucher force. Indeed, the formal difference is 
of fermions versus bosons in the loop. In section 2 we computed the long
range potential for very light pseudoscalars 
in the linear coupling scheme and also examined some
analogies and differences to the Feinberg--Sucher force. The latter included
some investigation on finite temperature corrections to the potentials.
The potential in this case falls off as $1/r^3$.
In the following section we performed a very similar exercise, but considering
a derivative coupling scheme for the interaction between heavy scalars
and pseudoscalars. Finally we presented a nice equivalence of both coupling
schemes in calculating the potential due to the exchange of true Goldstone
bosons. Here the fall--off is much steeper, namely $1/r^7$. As far as the
latter is concerned we add that a $1/r^5$ dependence is possible, via
box--diagrams, provided the pseudoscalars have tree level couplings to
fermions.

\acknowledgments
Work partially supported by the CICYT Research Project
AEN98-1093. F.F. acknowledges the CIRIT for financial support.
M.N. would like to thank Funda\c{c}\~ao de Amparo \`a Pesquisa de 
S\~ao Paulo (FAPESP) and Programa de Apoio a N\'ucleos de Excel\^encia
(PRONEX). 

\appendix
\section*{}

We present below the simplest version of a Majoron model which is a physical 
Goldstone boson in the spectrum of the theory associated with spontaneous
breakdown of total lepton number $L$ \cite{singlet}. 
This model, known as a singlet Majoron
model, became well known in connection with invisible Higgs decays
\cite{majoron1}. 
We 
emphasize that although the details will be given here for this particular 
model, a variety of similar models exist.

The usual motivation behind a Majoron model lies in the choice of the
Majorana mass term. The latter can be either a bare mass term,
$m_M\nu^T_R C \nu_R$, violating explicitly the lepton number or an interaction
term of the form $h\varphi \nu^T_R C\nu_R$ which conserves $L$. The field
$\varphi$ is a $SU(2)\otimes U(1)$ complex singlet with $L=-2$ which acquires
a non--zero vacuum expection value $<\varphi>=w/\sqrt{2}$ giving rise to a
Majorana mass term (h is a dimensionless parameter).

The scalar potential $V(\Phi, \varphi)$ contains besides the standard
Higgs doublet $\Phi$ the singlet $\varphi$. The potential is of the form
\begin{eqnarray} \label{potential}
V(\Phi, \varphi)&=& \mu^2_1(\Phi^{\dagger}\Phi) + \mu^2_2(\varphi^*\varphi)
+\lambda_1(\Phi^{\dagger}\Phi)^2 \nonumber \\
&+& \lambda_2(\varphi^*\varphi)^2 + \lambda_{12}(\Phi^{\dagger}\Phi)
(\varphi^* \varphi)
\end{eqnarray}
such that it conserves the lepton number.
We choose first the linear represention for the fields
\begin{equation} \label{rep1}
\displaystyle{\Phi =\left(\begin{array}{c}
G^+ \\
\displaystyle{
{v \over \sqrt{2}}+{\phi +iG^0 \over \sqrt{2}}} \end{array}\right), \,\,\,\,
\varphi= {w \over \sqrt{2}} + {\sigma +iJ \over \sqrt{2}}}
\end{equation}
where $G^+$ and $G^0$ are non--physical Goldstone bosons eaten up by the gauge
bosons according to the Higgs mechanism, $J$ is the 
the physical one (Majoron) and $v$ and $w$ are the corresponding vacuum 
expectation values triggering e.w. and lepton number S.S.B.. After minimization
of the potential the mass matrix of the two scalar particles reads
\begin{equation} \label{massmatrix}
\displaystyle{\left(\;\phi\;\;\; \sigma \; \right)
\left(\begin{array}{cc}
\lambda_1v^2 & \displaystyle{{\lambda_{12} \over 2}} vw \\
\displaystyle{{\lambda_{12} \over 2}} vw & \lambda_2 w^2 \end{array}\right)
\left(\begin{array}{c}
\phi \\
\sigma \end{array}\right)
={1 \over 2} m_H^2 HH + {1 \over 2}m^2_S SS}
\end{equation}
where $H$ and $S$ are the masss eigenstates obtained by the rotation
\begin{equation} \label{rotation}
\left(\begin{array}{c}
H \\
S \end{array}\right)=\left(\begin{array}{cc}
\cos \theta & -\sin \theta \\
\sin \theta & \cos \theta \end{array}\right)
\left(\begin{array}{c}
\phi \\
\sigma \end{array}\right).
\end{equation}
Equations (\ref{massmatrix}) and (\ref{rotation}) can be combined to deduce 
the following set of equations
\begin{eqnarray} \label{set}
2\lambda_1 v^2&=& \cos^2 \theta m^2_H + \sin^2 \theta m_S^2
\nonumber \\
2\lambda_2 w^2&=& \sin^2 \theta m^2_H + \cos^2 \theta m_S^2
\nonumber \\
2\lambda_{12} vw&=& \sin 2\theta (m^2_S - m_H^2).
\end{eqnarray}
Equation (\ref{set}) is useful to extract the vertices in terms
of the angle $\theta$ and the scalar masses. We are especially 
interested here in the trilinear vertices $HJ^2$ and $SJ^2$. They are given by
the interaction lagrangian
\begin{equation} \label{int1}
{\cal L}^{(1)}_{int}={(\sqrt{2} G_F)^{1/2} \over 2} 
\tan \beta \left[m^2_S \cos \theta S- m^2_H \sin \theta H \right]J^2+ ...
\end{equation}
where $G_F$ is the Fermi coupling constant and $\tan \beta =v/w$.

For comparison, let us also make use of a non--linear representation for
the singlet field $\varphi$, viz
\begin{equation} \label{rep2}
\varphi ={1 \over \sqrt{2}} (w + \sigma')\exp{(iJ/w)}.
\end{equation}
The components $\phi$ and $\sigma'$ will now mix to give the physical
scalars $H$ and $S$ (as in (\ref{massmatrix}) and (\ref{rotation})).
So far, there is no difference with respect to the linear representation. However,
in the non--linear representation the interaction terms of the Majoron
$J$ with the scalars will get generated in the singlet kinetic term
$(\partial_{\mu}\varphi^*)(\partial^{\mu}\varphi)$ which after rotation
to the physical scalars gives
\begin{equation} \label{int2}
{\cal L}^{(2)}_{int}=(\sqrt{2}G_F)^{1/2}\tan \beta \left[
\cos \theta S - \sin \theta H \right ]
(\partial_{\mu}J)(\partial^{\mu}J)+...
\end{equation}

As mentioned before, there exist a wide class of different Majoron models
invoking slightly different $U(1)$ symmetries to be spontaneously broken.
The latter can be either the lepton number, 
a combination of individual lepton numbers or a family symmetry. We refer
the reader to \cite{majoron2} 
for a short account of these models and references.
We mention also that some, previously popular Majoron models , like the
triplet--model or the doublet--model have been, by now, excluded in their
simplest versions through LEP data (through the absence of the decay
channel $Z \to J +{\mathrm Higgs}$). However, more complicated version
(mostly in conjunction with a singlet) can be still viable.

Also note that Majoron models which predict a tree level coupling to
ordinary matter are severely constrained by the argument of energy loss
in stars possibly carried away by Majorons. A singlet Majoron model evades
these constraints.

\end{document}